\documentclass[prd,twocolumn,aps,psfig,nofootinbib,nobibnotes,superscriptaddress,preprintnumbers,times]{revtex4-2}

\usepackage{graphicx,bm,color,amsmath,amssymb,mathtools,siunitx}
\usepackage{enumerate}

\graphicspath{ {./images/} }

\usepackage[dvipsnames]{xcolor}

\usepackage[hidelinks]{hyperref}
\hypersetup{
  colorlinks   = true, 
  urlcolor     = [RGB]{52,152,219}, 
  linkcolor    = [RGB]{207,54,108}, 
  citecolor    = [RGB]{52,152,219} 
}

\newcommand{\Mpc}{\text{Mpc}}
\newcommand{\km}{\text{km}}
\newcommand{\s}{\text{s}}
\usepackage{orcidlink}

\begin{document}
\title{Connecting Inflation to the NANOGrav 15-year Data Set via Massive Gravity}
\date{\today}
\author{Ved Kenjale\,\orcidlink{0009-0007-7186-9247}}
\email{vedk@cmu.edu}
\affiliation{McWilliams Center for Cosmology and Astrophysics and Department of Physics, \href{https://ror.org/05x2bcf33}{Carnegie Mellon University}, Pittsburgh, Pennsylvania 15213, USA}

\author{Tina~Kahniashvili\,\orcidlink{0000-0003-0217-9852}}
\email{tinatin@andrew.cmu.edu}
\affiliation{McWilliams Center for Cosmology and Astrophysics and Department of Physics, \href{https://ror.org/05x2bcf33}{Carnegie Mellon University}, Pittsburgh, Pennsylvania 15213, USA}
\affiliation{School of Natural Sciences and Medicine, \href{https://ror.org/051qn8h41}{Ilia State University}, 0194 Tbilisi, Georgia}
\affiliation{\href{https://ror.org/02gkgrd84}{Abastumani Astrophysical Observatory}, Tbilisi GE-0179, Georgia}

\begin{abstract}
Several pulsar timing array (PTA) missions have reported convincing evidence of a stochastic gravitational wave background within their latest datasets. This background could originate from an astrophysical source, though there are multiple possibilities for its origin to be cosmological. Focusing on the NANOGrav signal, which was in good agreement with other PTAs, we evaluate the possibility of an inflationary source for the background. 
However, we'll consider a time-dependent minimal theory of massive gravity instead of standard general relativity for our analysis. 
We find that an inflationary interpretation will require a strongly blue spectrum, characterized by $n_T = 1.8 \, \pm \, 0.54$, while Big Bang Nucleosynthesis limits will require a low reheating scale of $T_\text{rh} \simeq 4000 \ \text{GeV}$. Although these constraints make it difficult for inflation to be the source of the NANOGrav signal, we find that the massive gravity model allows for a greater reheating temperature than standard general relativity, making an inflationary interpretation slightly less cumbersome. 
\end{abstract}

\maketitle
\section{Introduction}
Inflation has become the widely accepted model for describing the evolution of the early universe \cite{Starobinsky:1979ty,Guth:1980zm,Linde:1981mu}. The inflationary paradigm solves the horizon, flatness, and magnetic monopole (non-existence) problems, and produces a stochastic background of relic gravitational waves (GWs) via parametric amplification of zero-point quantum-mechanical fluctuations in the gravitational field \cite{Grishchuk:1974ny,Starobinsky:1979ty} (see \cite{Maggiore:v2} for a review). There has yet to be a detection of relic GWs, but such an event could {  (in)validate} inflation, test { gravity} and other fundamental physics, and provide unique insight into the characteristics of the very early universe prior to recombination \cite{Lidsey:1995np,Maggiore:1999vm,Braglia:2020taf,Roshan:2024qnv}. 
An avenue for the detection of primordial GWs lies in pulsar timing arrays (PTAs), which utilize millisecond pulsars as stable clocks \cite{Verbiest_2009}\footnote{{ A promising competitive detection method involves astronometry; see \cite{Qin:2018yhy} for a review and \cite{Moore:2017ity, Mihaylov:2018uqm, Mihaylov:2019lft, Liang:2023pbj, Inomata:2024kzr} for original studies exploring the synergy between astrometry and PTAs.}}. GWs perturb the spacetime along the line of sight to a pulsar, causing spatially correlated fluctuations \cite{Bernardo:2023mxc,2024arXiv240907955B} to manifest themselves in the arrival time of pulses \cite{Sazhin:1978,Detweiler:1979wn}.
{ As such, comparing theoretically predicted GWs with observations places constraints on new physics operating at frequencies within the sensitivity range of pulsar timing arrays  \cite{NANOGrav:2023hvm}.}

Recently, the North American Nanohertz Observatory for Gravitational Waves (NANOGrav) \cite{Agazie:2023}, the European Pulsar Timing Array (EPTA) \cite{EPTA:2023sfo, EPTA:2023fyk, EPTA:2023akd}, the Parkes Pulsar Timing Array (PPTA) \cite{Zic:2023gta, Reardon:2023gzh}, and the Chinese Pulsar Timing Array (CPTA) \cite{Xu:2023wog} collaborations released analyses of their newest PTA datasets \cite{InternationalPulsarTimingArray:2023mzf}. Their analyses show evidence of Hellings-Downs correlations \cite{Hellings:1983fr}, which indicate a true GW source for the observed PTA signal and support the existence of a stochastic GW background \cite{Romano:2016dpx}. These datasets will also allow us to test fundamental physics \cite{NANOGrav:2023hvm, Agazie:2024kdi}), including general relativity (GR) (see Ref. \cite{Yunes:2013dva} for a living review, and for recent works \cite{Liang:2021bct, Liang:2023ary,  Bernardo:2023pwt, Bernardo:2023jhs, Wu:2023hsa, Hu:2024wub, Liang:2024mex, Bernardo:2024uiq, Hu:2024wub, Zheng:2024tib}). For the remainder of this work, we will consider the NANOGrav 15-year data set \cite{NANOGrav:2023gor}.

The detection of a stochastic GW background via PTAs could be understood as a signal from the early universe \cite{Bernardo:2024bdc}, although a component of this signal may arise from astrophysical sources \cite{Bian:2023dnv,NANOGrav:2023hvm}. Previous works \cite{Vagnozzi:2020gtf, Sakharov:2021dim, Lazarides:2021uxv,Brandenburg:2021pdv,   Vagnozzi:2023lwo, Odintsov:2023weg, Choudhury:2023hfm, Choudhury:2023kam, Niu:2023bsr,Datta:2023vbs, Jiang:2023gfe,Figueroa:2023zhu,Borah:2023sbc,Lazarides:2023rqf,Fu:2023aab,Correa:2023whf,Chen:2024mwg, Firouzjahi:2023lzg,Datta:2023xpr, Benetti:2021uea, Liu:2023pau, Liu:2023ymk, Datta:2023vbs, Datta:2023xpr} have investigated an interpretation of the NANOGrav signals
(the NANOGrav 15-year dataset \cite{NANOGrav:2023gor} and/or the previous NANOGrav 12.5-year dataset \cite{NANOGrav:2020bcs}) assuming GWs are 
generated in the early stages of the Universe expansion, in particular by inflation. 
These works found that a strongly blue spectrum and low reheating temperature would be needed to interpret the PTA signal as inflationary GWs \footnote{{ Cosmic Microwave Background (CMB) based relic GW tests give us a direct probe of inflation \cite{Starobinsky:1979ty}. In fact, in the standard cosmological scenario, the typical frequency of GWs from slow-roll inflation at current time is $10^{-20}-10^{-18}$ Hz, while their amplitude depends greatly on the inflationary model considered \cite{Caprini:2018mtu}. Though these GWs cannot be detected directly, they can be analyzed through the CMB temperature and polarization anisotropies (for pioneering works, see Refs. \cite{1994LNP...429..129S,Frewing:1993dq,Kosowsky:1994cy,
Polnarev:1995gk,Jungman:1995av, Jungman:1995bz})  
Specifically, the presence of inflationary GWs with substantial amplitude will manifest themselves through the induced B-mode of polarization (see Refs. \cite{Zaldarriaga:1996xe,Seljak:1996gy,Kamionkowski:1996zd,Kamionkowski:1996ks} and \cite{BICEP:2021xfz,LiteBIRD:2023aov,LiteBIRD:2024dbi} for recent constraints). 
Both inflationary tensor and scalar perturbations have the same source: quantum mechanical fluctuations \cite{Mukhanov:1990me}. Consequently, the spectra of scalar and tensor modes are interdependent, making the tensor-to-scalar ratio $r$ a parameter that depends on the specific inflationary model \cite{Lasky:2015lej}.  Therefore, inflationary models and their characteristics may be constrained through the CMB \cite{Planck:2018jri}. Although anisotropy maps serve as a probe of the physics of inflation, there is still a large degree of flexibility in the exact model building \cite{Martin:2013tda}. }}; hence, inflationary models beyond the simplest ones would be required. One possibility for modifying the inflationary paradigm is to relax the assumption of GR as the true theory of gravity (at all energy and length scales). 
Our goal is to determine whether modifications to GR would result in a more feasible interpretation of inflationary GW as a source of PTA signals. In particular, we will focus on a theory known as massive gravity (MG), where the graviton has a non-zero mass $m$ \cite{deRham:2016nuf}. 

The theory of MG was pioneered by Fierz and Pauli in 1939, since they considered the addition of a Lorentz invariant mass term to a linearized spin-2 field \cite{Fierz:1939ix}. MG has become quite relevant during the last few decades due to the accelerated expansion of the universe today, which can be explained by a graviton mass term without the need for dark energy or the presence of a cosmological constant \cite{DAmico:2011eto}. 
For quite some time, modifying the nil graviton mass proved to be difficult. Linear MG theories fail to reduce to GR in the massless graviton limit, due to the van Dam-Veltman-Zakharov (DVZ) discontinuity \cite{Zakharov:1970cc, vanDam:1970vg}.
\footnote{
MG theories have five polarization modes: in addition to the regular tensor (spin $\pm 2$) modes, there are scalar (spin 0) and vector (spin $\pm 1$) modes that do not disappear in the $m \rightarrow 0$ limit, i.e. regular GR is not recovered.}
Non-linear theories have also been considered: the non-linear extension to Fierz-Pauli MG is subject to the Vainshtein screening mechanism \cite{Vainshtein:1972sx}, which removes additional polarization modes. However, the theory still suffers from ghost degrees of freedom: a helicity-0 mode in the gravity sector may be absent at the quadratic order of Fierz-Pauli MG, but revive at a higher order, acting as a ghost \cite{Boulware:1972yco}. 

A more recent effort resulted in the breakthrough formulation of a ghost-free MG model: deRham-Gabadadze-Tolley (dRGT) \cite{deRham:2010ik,deRham:2010kj}, and its bi-gravity generalization \cite{Hassan:2011zr}. These formulations of viable MG models have led to investigations on the effect of a nonzero graviton mass on GW generation and propagation; see Refs. \cite{Gumrukcuoglu:2011zh,Gumrukcuoglu:2012wt} for pioneering works and Ref. \cite{deRham:2014zqa} for a review. 
Another option to avoid the Boulware-Deser ghost is to relax the need for Lorentz invariance \cite{Rubakov:2004eb,Blas:2007zz, Rubakov:2008nh, Blas:2009my}. Then, the massive graviton forms a representation of a 3D rotation group, instead of the 4D Lorentz group it would form with Lorentz invariance. As a result of this, the number of degrees of freedom in the gravity sector is no longer forced to be five. 

The minimal theory of MG (MTMG) is a model that utilizes this, retaining only two degrees of freedom in the gravity sector, from the two tensor modes \cite{DeFelice:2015hla, DeFelice:2015moy}. Lorentz-violating theories, like MTMG, also avoid the Higuchi bound, which requires that the Lorentz-invariant graviton mass should be greater than the Hubble expansion rate by a factor up to order unity to avoid turning extra degrees of freedom into ghosts in cosmological backgrounds \cite{Higuchi:1986py}. 

This work will utilize a time-dependent MTMG, with the graviton mass following a step function as in Ref. \cite{Fujita:2018ehq}. We will evaluate how MTMG can affect the inflationary interpretation of the NANOGrav 15-year dataset. The paper is organized as follows: In Section \ref{model setup}, we review the models for inflationary GWs, MG, and PTA data. In Section \ref{transfer function}, we connect PTA data and inflation using transfer functions. In Section \ref{data and methods}, we map our results to the inflationary regime and explore the implications of this in Section \ref{results}. We conclude and discuss future extensions in Section \ref{conclusions}. 
We will use natural units with $c = \hbar = k_B = 1$. We set the present-day Hubble parameter to $H_0 = 100 h_0\, \km \, \s^{-1} \, \Mpc^{-1} = 67.66 \, \km \, \s^{-1} \, \Mpc^{-1}$, noting that this value isn't a universal constant ({ i.e. 
it has large uncertainties}) due to the Hubble tension \cite{Planck:2018vyg}. We set the current radiation, matter and dark energy density parameters to $\Omega_R, \Omega_M, \Omega_\Lambda = \num{9.182e-5}, 0.3111, 0.6889$ using \textit{Planck} 2018 data \cite{Planck:2018vyg}. 

\section{Model Setup} \label{model setup}
We begin by defining the line element of the Friedmann–Lemaître–Robertson–Walker (FLRW) flat metric:
\begin{equation} \label{eq:metric}
    ds^2 = a^2(\tau) \left[-d\tau^2 + (\delta_{ij}+h_{ij})dx^idx^j \right] \ ,
\end{equation}
where $a(\tau)$ and $\tau$ denote the scale factor and conformal time respectively, and $h_{ij}$ denotes the traceless and transverse (i.e. $g^{ij}h_{ij} = \partial_i h_{ij} = 0$) tensor perturbation to flat space. 
Going into momentum space, we decompose the tensor perturbation $h_{ij}$ into its helicity states (see Eq. 19.214 of \cite{Maggiore:v2}):
\begin{equation} \label{eq: helicity decomp}
    h_{ij}(\tau, \boldsymbol{k}) = \sum_{\lambda = +, \times} e_{ij}^\lambda (\boldsymbol{k}) h_k^\lambda(\tau, \boldsymbol{k}) \ ,
\end{equation}
where $\boldsymbol{k}$ is the comoving momentum, $+$ and $\times$ denote the two polarization states, and $e_{ij}^\lambda$ are the polarization tensors defined by (see Eqs. 19.216-217 of \cite{Maggiore:v2}): 
\begin{align}
      e^+_{ij}(\hat{\boldsymbol{k}}) &= \hat{\boldsymbol{u}}_i\hat{\boldsymbol{u}}_j - \hat{\boldsymbol{v}}_i\hat{\boldsymbol{v}}_j 
    \\e^\times_{ij}(\hat{\boldsymbol{k}}) &= \hat{\boldsymbol{u}}_i\hat{\boldsymbol{v}}_j - \hat{\boldsymbol{v}}_i\hat{\boldsymbol{u}}_j \ .
\end{align}
The unit vectors $\hat{\boldsymbol{u}}$ and $\hat{\boldsymbol{v}}$ are orthogonal to each other and $\hat{\boldsymbol{k}}$, and the polarization tensors are also normalized as: 
\begin{equation}
    e_{ij}^\lambda (\hat{\boldsymbol{k}}) e_{ij}^{\lambda'} (\hat{\boldsymbol{k}}) = 2\delta^{\lambda \lambda'} \ .
\end{equation}
We now introduce the relevant equations for MTMG. First, the general quadratic action is given by \cite{Gumrukcuoglu:2012wt, Fujita:2018ehq, DeFelice:2015hla, DeFelice:2015moy, Choi:2023tun}:
\begin{equation} \label{eq: action}
    S = \frac{M_\text{Pl}^2}{8} \int d\tau d^3x \ a^2 \Bigl [ (h_{ij}')^2 - (\partial h_\text{ij})^2 - a^2 M_\text{gw}^2 h_{ij}^2 \Bigr ] \ ,
\end{equation}
where $M_\text{Pl}$ is the Planck mass, $M_\text{gw}$ is the graviton mass
\footnote{ 
The graviton mass and the mass of GWs may differ in some scenarios, see Refs. \cite{DAmico:2012hia, Kahniashvili:2014wua, DeFelice:2013tsa, DeFelice:2013dua, DAmico:2013saf, Gannouji:2013rwa, Mukohyama:2014rca} for examples using a quasidilaton theory of MG.
},
and the primes ($'$) denote derivatives with respect to the conformal time $\tau$. The graviton mass 
will follow a step function, given by:
\begin{equation} \label{eq: graviton mass}
    M_\text{gw}(\tau) = 
    \begin{cases}
        m & \tau < \tau_m \\
        0 & \tau > \tau_m
   \end{cases} \ ,
\end{equation}
{where $m$ is a fixed (see Section \ref{transfer function}) non-zero mass value, and $\tau_m$, the cutoff time, denotes the conformal time at which the graviton mass drops to nil.} Although this instantaneous change to zero in the graviton mass is somewhat unphysical, specifying an exact form of this function would require fixing a model that incorporates the dynamics of other fields. However, the overarching effect of a time-dependent theory of MG can still be studied using the simple step function mass, taking advantage of the simplified analytical and numerical calculations permitted. Since we will also consider GR, we note that $M_\text{gw}(\tau) = 0$ for GR. 

Following the standard procedure of 
minimizing the action, Eq. \ref{eq: action}, and utilizing the helicity decomposition from Eq. \ref{eq: helicity decomp}, we obtain the equation of motion (EoM) for $h_k$ \cite{Gumrukcuoglu:2012wt, Fujita:2018ehq}:
\begin{equation} \label{eq: EoM}
    \bar{h}_k'' + \left (k^2 + a^2M_\text{gw}^2 - \frac{a''}{a} \right)\bar{h}_k = 0 \ ,
\end{equation}
where $\bar{h}_k = a \cdot h_k$, and we've suppressed the subscript $\lambda$ on $h_k^\lambda$ denoting the polarization mode since the two polarization modes follow the same EoM (i.e. unpolarized GWs). From the EoM, we also acquire the dispersion relation:
\begin{equation}
    \omega^2 = \frac{k^2}{a^2} + M_\text{gw}^2 \ ,
\label{eq: dispersion}
\end{equation}
which relates the frequency $\omega$ to the wavenumber $k$. 

From an initial conformal time $\tau_i$, a mode $h_k$ will evolve according to Eq. \ref{eq: EoM}, which we summarize via the transfer function:
\begin{equation} \label{eq: transfer func}
    \mathcal{T}(\tau, k) = \frac{h_k(\tau)}{h_k(\tau_i)} \ .
\end{equation}
To characterize the GW signal, 
we define the tensor power spectrum as: 
\begin{equation}
    \mathcal{P}_T(\tau, k) \equiv \frac{k^3}{\pi^2} \sum_{\lambda = +, \times}  \vert h_k^\lambda(\tau)\vert ^2 = \frac{2k^3}{\pi^2} \vert h_k(\tau)\vert ^2 \ .
\end{equation}
We can recast this in terms of the transfer function by first defining the primordial tensor power spectrum:
$\mathcal{P}_T^\text{prim}(k) \equiv \mathcal{P}_T(\tau_i, k)$. It follows that:
\begin{equation}
    \mathcal{P}_T(\tau, k) = \mathcal{P}_T^\text{prim}(k) \mathcal{T}^2(\tau, k) \ .
\end{equation}
The relevant quantity for GW detection experiments is the GW spectral energy density, $\Omega_\text{gw}$, a dimensionless quantity that describes the strength of GWs in terms of their energy density. It is given by the logarithmic derivative with respect to wavenumber of the GW energy density $\rho_\text{gw}$, divided by the critical density $\rho_c$ \cite{Maggiore:1999vm}, and is evaluated at a conformal time $\tau$:
\begin{equation} \label{eq: GWSED def}
    \Omega_\text{gw}(\tau, k) \equiv \frac{1}{\rho_c} \frac{d \rho_\text{gw}}{d\ln k} \ .
\end{equation}
This can be defined in terms of the transfer function by first identifying $\rho_\text{gw}$ as the 00-component of the energy-momentum tensor $T^{\mu \nu}$ \cite{Kuroyanagi:2008ye}:
\begin{align} \label{eq: rho_gw}
    T^{00} = \rho_\text{gw} &= \frac{1}{64\pi G a^2} \langle (\partial_\tau h_{ij})^2 + (\vec \nabla h_{ij})^2 \rangle \\
    &= \frac{1}{32 \pi G} \int \frac{d^3k}{(2 \pi)^3} \frac{k^2}{a^2} \cdot 2\sum_{\lambda = +, \times} \vert h_k^\lambda \vert^2 \ ,
\end{align}
Substituting this into Eq. \ref{eq: GWSED def} yields the desired form \cite{Kuroyanagi:2010mm, Kuroyanagi:2011fy}:
\begin{equation} \label{eq: GWSED transfer}
    \Omega_\text{gw}(k) = \frac{k^2}{12 H_0} \mathcal{T}^2(\tau_0, k) \mathcal{P}_T^\text{prim}(k) \ ,
\end{equation}
where we've evaluated the expression at $\tau = \tau_0$. This form of the GW spectral energy density is useful since it contains two distinct factors: the transfer function, $\mathcal{T}(\tau_0, k)$ and the primordial tensor power spectrum $\mathcal{P}_T^\text{prim}(k)$. Let's set the initial conformal time $\tau_i$ from which the transfer function is defined by to the conformal time at which inflation ends. Then, $\mathcal{P}_T^\text{prim}(k)$ gives the tensor power spectrum at the end of inflation. This also means that $\mathcal{T}(\tau_0, k)$ represents the evolution of a GW mode in the universe from the end of inflation till the present time. Hence, this definition of $\tau_i$ has effectively split the GW spectral energy density into a part dependent on inflation, and by extension a particular inflationary model, and a part independent of inflation. This latter part, $\mathcal{T}(\tau_0, k)$, is governed by Eq. \ref{eq: EoM} and thus depends on the model of gravity being considered. 

The specific form of the transfer function $\mathcal{T}(\tau_0, k)$ will be discussed later, but to connect to inflation, we must specify the form of the primordial tensor power spectrum $\mathcal{P}_T^\text{prim}(k)$. Since we will eventually constrain the inflationary period via the NANOGrav signal, we approximate the primordial tensor power spectrum as a power-law, as is convention \cite{Maggiore:v2}:
\begin{equation} \label{eq: prim power spectrum}
    \mathcal{P}_T^\text{prim}(k) = r A_s \left ( \frac{k}{k_\star} \right )^{n_T} \ ,
\end{equation}
where 
$A_s$ is the amplitude of primordial scalar perturbations at the CMB pivot scale $k_\star$, chosen as $k_\star = 0.05 \ \Mpc^{-1}$ in this work to match \textit{Planck} 2018 \cite{Planck:2018vyg}, and $n_T$ is the tensor spectral index. If $n_T = 0$, the spectrum is scale-invariant (no dependence on $k$), while $n_T < 0$ is called as a red spectrum and $n_T > 0$ is called a blue spectrum. The single-field slow-roll inflationary model adheres to a ``consistency relation" given by $n_T = -r / 8$ \cite{Copeland:1993ie}, which would imply a red spectrum for this model given that $r > 0$. The spectral index $n_T$ is also related to the equation-of-state $w$ for the dominant component during the inflationary period by \cite{Maggiore:v2}:
\begin{equation} \label{eq: eos n_T}
    n_T = \frac{4}{1 + 3w} + 2 \ .
\end{equation}
An inflationary period characterized by de Sitter expansion would have $w = -1$, and thus $n_T = 0$ for a scale-invariant spectrum. For simple scalar-field inflationary models (including single-field slow-roll inflation), we have $w > -1$, yielding a slightly red spectrum with $n_T < 0$ \cite{Linde:1981mu}. For phantom inflationary models, we have $w < -1$ and thus $n_T > 0$, i.e. a blue spectrum \cite{Piao:2004tq}. It is also important to note that the constraint $w < -1/3$ (see Section 17.2 of Ref. \cite{Maggiore:v2}), needed for the universe to follow an accelerating expansion during inflation, implies that $n_T < 2$. 

As mentioned previously, we will utilize the GW spectral energy density to connect our model to an observed signal (i.e. NANOGrav 15-year dataset). For a PTA signal, this is typically given as:
\begin{equation} \label{eq: GWSED PTA}
    \Omega_\text{gw}^\text{PTA}(f) = \frac{2 \pi^2}{3 H_0^2}f^2 h_c^2(f) \ ,
\end{equation}
where $h_c(f)$ is the power spectrum of the GW strain, approximated as the power law:
\begin{equation} \label{eq: GW strain}
    h_c(f) = A \left ( \frac{f}{f_\text{yr}} \right )^\alpha \ ,
\end{equation}
where $A$ is the intrinsic-noise amplitude, the reference frequency $f_\text{yr} = 1 \ \text{yr}^{-1}$, and the spectral index $\alpha$ is given by:
\begin{equation} \label{eq: alpha, gamma}
    \alpha = \frac{3 - \gamma}{2} \ ,
\end{equation}
where $\gamma$ is spectral index for the pulsar timing residual cross-power spectral density $S(f) \propto f^{-\gamma}$ \cite{NANOGrav:2023gor}. Using Eqs. \ref{eq: GWSED PTA}, \ref{eq: GW strain}, and \ref{eq: alpha, gamma} we can express the GW spectral energy density for a PTA signal as:
\begin{equation}
    \Omega_\text{gw}^\text{PTA}(f) = A^2 \frac{2 \pi^2}{3 H_0^2} \frac{f^{5 - \gamma}}{\text{yr}^{\gamma - 3}} \ .
\end{equation}
The intrinsic-noise amplitude $A$ and the spectral index $\gamma$ are used to characterize the results of PTA GW experiments. The NANOGrav collaboration reports their results as joint $A$-$\gamma$ posteriors or posteriors for $A$ at a fixed value of $\gamma$, often $\gamma = 13/3$ (the expected value of $\gamma$ for a stochastic GW background originating from supermassive black hole binaries (SMBHBs))
\cite{NANOGrav:2023gor}. We shall utilize the joint $A$-$\gamma$ posteriors in Section \ref{data and methods}. 

\section{Transfer Functions} \label{transfer function}
In this section, we work towards numerically integrating the EoM (Eq. \ref{eq: EoM}) to calculate the transfer function $\mathcal{T}(\tau_0, k)$ (Eq. \ref{eq: transfer func}) in the GR and MG paradigms. These transfer functions will then be used to calculate the GW spectral energy density (Eq. \ref{eq: GWSED transfer}) before connecting to PTA data.  

The form of EoM requires that we specify the graviton mass $M_\text{gw}$ and the scale factor $a$ as functions of conformal time, over the relevant time domain. In our case, we will integrate (over conformal time) from the end of inflation to the present day, and we'll use $\tau_0 = \num{1.375e4} \ \Mpc$ (see. Eq. 17.168 of \cite{Maggiore:v2}). The functional form for the graviton mass was given in Eq. \ref{eq: graviton mass}, but the non-zero mass $m$ and the cutoff time $\tau_m$ must be specified. { We now discuss some of the effects that can phenomenologically be used to constrain the graviton mass (see Ref. \cite{deRham:2016nuf} for a review).
\begin{enumerate}[i.]
    \item {\bf Propagation Speed and Dispersion Relation:} 
    Changes in the dispersion relation (Eq. \ref{eq: dispersion}), namely a difference in the GW propagation speed from the speed of light, result in a model-dependent upper bound of the graviton mass.  
    The first and third GW Transient Catalogs (GWTC-1 and GWTC-3) enabled the search for cosmic strings and can probe the graviton mass at scales around $10^{-23}$ eV \cite{LIGOScientific:2021nrg,LIGOScientific:2021sio}. The analysis of the PTA dataset has restricted this to about $3.8 \times 10^{-23}$ eV \cite{Wu:2023pbt,Wang:2023div}.
    CMB and baryon acoustic oscillations further constrain the graviton mass in extended MTMG to around $6.6 \times 10^{-34}$ eV \cite{DeFelice:2022mcd, DeFelice:2023bwq}.
    
    \item {\bf  Yukawa's Suppression on the Gravitational Potential:}
    The presence of a non-zero graviton mass results in a Yukawa-type exponential suppression of the gravitational potential. This could lead to limits of the order of $10^{-29}$ eV from observations of gravitationally bound clusters of 0.5 Mpc
    \cite{Goldhaber:1974wg} (see also the Particle Data Group's limits \cite{ParticleDataGroup:2010dbb}).
    
    \item {\bf Additional Polarization Modes and Fifth Force. }
    The presence of additional helicity-$0$ and helicity-$\pm 1$ modes modifies the Hellings-Downs curve through induced monopole and dipole contributions \cite{NANOGrav:2021ini,NANOGrav:2023ygs}. Indeed, the detection of the additional GW polarization modes in PTA data is a promising way to test beyond the standard model scenarios \cite{Bernardo:2022rif}. 
    The main consequence of additional degrees of polarization for the graviton is a coupling of the helicity-$0$  mode to matter (the so-called {\it fifth force}) \cite{deRham:2014naa}. However, in dense environments,
    the Vainshtein mechanism \cite{Vainshtein:1972sx} ensures the screening of the helicity-$0$ (and helicity-$\pm 1$) mode \cite{deRham:2012fw,Falck:2014jwa,Bloomfield:2014zfa,Falck:2015rsa,
    Koyama:2015oma,Kase:2015zva}, while leaving the helicity-$\pm 2$ modes the same as in GR.
    Strong suppression of the helicity-$0$ mode in MG models makes its observation via interferometer GW missions problematic, but additional polarization degrees (one scalar helicity-$0$, two vector helicity-$\pm 1$, and two tensor helicity-$\pm 2$ modes \cite{Eardley:1973br,Eardley:1974nw}), can be seen and constrained by solar system, pulsar, black hole, and cosmology tests \cite{Hui:2012jb,deRham:2014zqa}. Additionally, 
    fifth force measurement based tests could (possibly) lead to limits of the order of $10^{-32}-10^{-33}$ eV through weak lensing from future surveys \cite{Zu:2013joa,Wyman:2011mp,Park:2014aga}.
    
\end{enumerate}
} 
{The present day graviton mass is also bounded by the relation $M_\text{gw} \lesssim \num{2e-28} \text{eV} / c^2$ from the orbital decay of binary pulsars \cite{Shao:2020fka, Finn:2001qi}. However, since we are considering a step function mass, it's not necessary to follow the constraints discussed above. We choose $m = 10^9 \cdot H_0$ to match Refs. \cite{Gumrukcuoglu:2012wt, Fujita:2018ehq}. This choice is somewhat arbitrary, but since we'll require a fixed value going forward, we accept it and understand that uncertainty in the graviton mass will be reflected onto the magnitude of the transfer functions calculated below. }

As for the cutoff time, the binary neutron star merger labeled as events GW170817 and GRB170817A occurred at $z \simeq 10^{-2}$ and bounded the propagation speed of gravitons \cite{LIGOScientific:2017zic}. However, we choose to conservatively assume that the cutoff time occurs around the matter-radiation equality. This choice will also simplify the numerical simulation, as only part of the matter dominated era will have a non-zero graviton mass. 
In particular, we choose:
\begin{align}
    \tau_m &= 5 \tau_\text{eq} \\
    &= 5 (\sqrt{2} - 1) \tau_\star \\
    &= 5 (\sqrt{2} - 1) \frac{2 \Omega_R^{1/2}}{H_0 \Omega_M} \ ,
\end{align}
where $\tau_\text{eq}$ is the conformal time at matter-radiation equality, $\tau_\star$ is a conformal time scale, and the last two lines come from Eqs. 17.153-154 of \cite{Maggiore:v2}. For the scale factor, we'll use:
\begin{equation}
    a(\tau) = a_\text{eq} \left [ \frac{2 \tau}{\tau_\star} + \left ( \frac{\tau}{\tau_\star} \right )^2 \right ] \ ,
\end{equation}
which holds during the radiation dominated and matter dominated epochs (see Eq. 17.152 of \cite{Maggiore:v2}). We do not explicitly consider a dark energy dominated era, but the effects of this era will be included as an additional factor in the transfer function. This factor is approximated as \cite{Zhao:2006mm, Kuroyanagi:2014nba}:
\begin{equation}
    \mathcal{T}_2 = \frac{\Omega_M}{\Omega_\Lambda} \simeq 0.451 \ ,
\end{equation}
using the values previously defined for $\Omega_M$ and $\Omega_\Lambda$. Another effect to note is the damping caused by free-streaming neutrinos from decoupling in the early universe. This free-streaming produces an anisotropic stress \cite{Durrer:1997ta}, which can reduce the amplitude of modes that enter the horizon during the radiation dominated era \cite{Weinberg:2003ur}. Defining $\mathcal{T}_3$ to be the factor associated with neutrino free-streaming, we have:
\begin{equation} \label{eq: TF all factors}
    \mathcal{T} = \mathcal{T}_1 \cdot \mathcal{T}_2 \cdot \mathcal{T}_3 \ ,
\end{equation} 
where $\mathcal{T}_1$ is the contribution from the numerical integration of the EoM. 
\begin{figure}[ht]
\centering
\includegraphics[width=\linewidth]{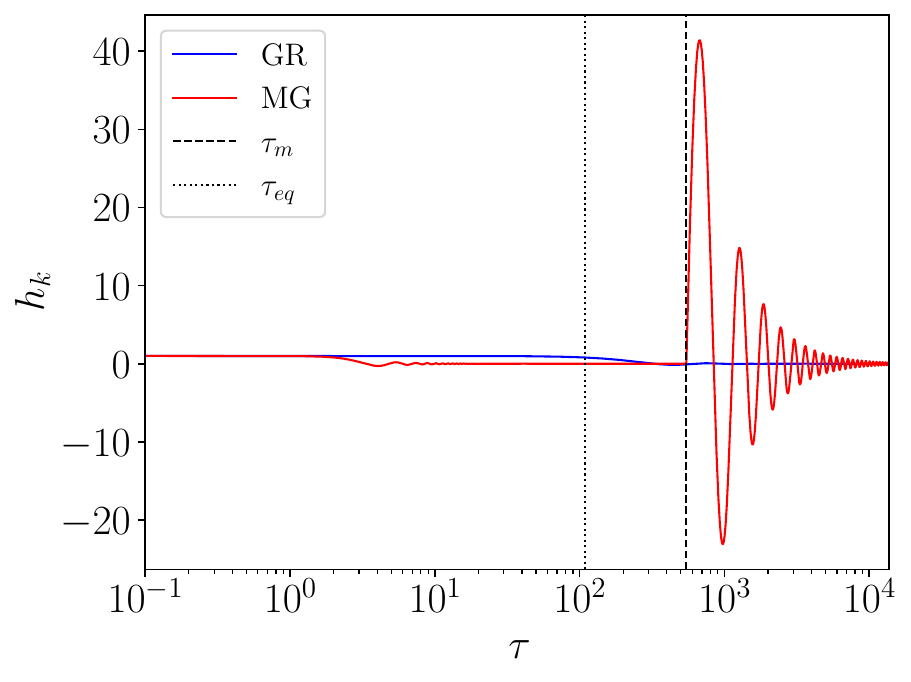}
\caption{Evolution of the mode $k = k_\text{eq}$ in the GR and MG models, where $k_\text{eq} = 0.010339 \ \Mpc^{-1}$ is the wavenumber for a mode crossing the horizon at matter-radiation equality \cite{Planck:2018vyg}. 
This simulation assumes $h_k = 1$ prior to the mode entering the horizon. Observe that in the GR case, the mode does enter the horizon at $\tau = \tau_\text{eq}$, where $\tau_\text{eq}$ is the conformal time of horizon reentry for $k_\text{eq}$ (note the logarithmic scale for $\tau$). }
 \label{fig: mode evolution}
\end{figure}

With the graviton mass $M_\text{gw}$ and the scale factor $a$ specified, we now proceed with the numerical integration of the EoM. The evolution of a single mode in the GR and MG paradigms is shown in Fig. \ref{fig: mode evolution}. We repeat this integration for a domain of $k$ values, allowing us to calculate the transfer functions in the GR and MG cases:
\begin{align}
    \mathcal{T}_1^{\text{GR}}(k) &= \frac{3 j_1(k \tau_0)}{k \tau_0} \sqrt{0.96 + 0.97 \left( \frac{k}{k_\text{eq}} \right) + 4.16 \left( \frac{k}{k_\text{eq}} \right)^2} \label{eq: GR TF}
    \\  \mathcal{T}_1^{\text{MG}}(k) &= \frac{3 j_1(k \tau_0)}{k \tau_0} \nonumber
    \\ & \cdot \sqrt{2.1\cdot10^4 + 5.8\cdot10^4 \left( \frac{k}{k_\text{eq}} \right) + 1.3 \cdot 10^6 \left( \frac{k}{k_\text{eq}} \right)^2}  \ ,\label{eq: MG TF}
\end{align}
where $j_1$ is the first spherical Bessel function, given by $j_1(x) = \sin(x) / x^2 - \cos(x) / x$. The GR transfer function has good agreement with previous works, which found similar coefficients \cite{Turner:1993vb, Kuroyanagi:2014nba}. Since the GR and MG case transfer functions have the same functional form (which was utilized to perform function fitting to obtain the coefficients shown in Eqs. \ref{eq: GR TF} and \ref{eq: MG TF}), we'll simply use:
\begin{equation} \label{eq: SK TF}
    \mathcal{T}_1(k) = \frac{3 j_1(k \tau_0)}{k \tau_0} \sqrt{s_3 + s_2 \left( \frac{k}{k_\text{eq}} \right) + s_1 \left( \frac{k}{k_\text{eq}} \right)^2} \ ,
\end{equation}
and insert the appropriate values of $s_1, s_2,$ and $s_3$ when necessary. Note that the NANOGrav PTA signal occurs at $f \sim 10^{-9} \ \text{Hz}$, which corresponds to $k \sim 10^6 \ \Mpc^{-1}$. Since this scale is much greater than $k_\text{eq}$, we can simplify Eq. \ref{eq: SK TF} by working in the $k \gg k_\text{eq}$ regime. By also using the envelope of the $j_1$ spherical Bessel function, i.e. $j_1(x) \simeq 1 / x$, we obtain:
\begin{equation} \label{eq: simplified TF}
     \mathcal{T}_1(k) \simeq \frac{3}{(k\tau_0)^2} \sqrt{s_1} \left ( \frac{k}{k_\text{eq}} \right ) \ .
\end{equation}
In this regime, the neutrino free-streaming factor, $\mathcal{T}_3$, is one (see Ref. \cite{Zhao:2006mm} for more details). 
As such, we express the complete transfer function as:
\begin{equation} \label{eq: full TF, squared}
    \mathcal{T}^2(k) = \frac{9 s_1}{k^2 \tau_0^4 k_\text{eq}^2} \frac{\Omega_M^2}{\Omega_\Lambda^2} \ ,
\end{equation}
where in the $k \gg k_\text{eq}$ regime, the only relevant coefficient from Eq. \ref{eq: SK TF} is $s_1$. Inserting $\mathcal{T}^2(k)$ from Eq. \ref{eq: full TF, squared} and $\mathcal{P}_T^\text{prim}(k)$ from Eq. \ref{eq: prim power spectrum} into Eq. \ref{eq: GWSED transfer} gives: 
\begin{equation} \label{eq: GWSED infl}
    \Omega_\text{gw}(k) = \frac{3 s_1}{4 \tau_0^4 H_0^2 k_\text{eq}^2} \frac{\Omega_M^2}{\Omega_\Lambda^2} \cdot r A_s \left ( \frac{k}{k_\star} \right )^{n_T} \ .
\end{equation}
We can now connect the inflationary and PTA parameters via the GW spectral energy density. In particular, we equate $\Omega_\text{gw}(k)$ from Eq. \ref{eq: GWSED infl} to $\Omega_\text{gw}^\text{PTA}(f)$ from Eq. \ref{eq: GWSED PTA}. This first yields a relation between the tensor spectral index $n_T$ and the spectral index $\gamma$:
\begin{equation} \label{eq: n_T(gamma)}
    n_T = 5 - \gamma \ .
\end{equation} 
Utilizing this relation and further equating the PTA and inflationary GW spectral energy densities gives a relation between the intrinsic-noise amplitude $A$ and the inflationary parameters $r$ and $n_T$:
\begin{equation} \label{eq: A(n_T, r)}
    A = \frac{\Omega_M}{\Omega_\Lambda} \sqrt{\frac{9 s_1 A_s \text{yr}^2}{8 \pi^2 \tau_0^4 k_\text{eq}^2}} \left ( \frac{\text{yr}^{-1}}{f_\star}\right )^{n_T / 2} \sqrt{r} \ ,
\end{equation}
which, along with Eq. \ref{eq: n_T(gamma)}, will be our primary connection between the NANOGrav 15-year posteriors and the inflationary parameter space. 

Since primordial GWs act like extra radiation as they propagate within the horizon, we must consider their contribution to the radiation energy density of the early universe. This contribution has effects on Big Bang Nucleosynthesis (BBN), which provides an important constraint \cite{Kuroyanagi:2014nba, Boyle:2007zx, Giare:2022wxq, Maggiore:1999vm, Cabass:2015jwe}:
\begin{equation} \label{eq: BBN integral}
    \Delta N_\text{eff}^\text{gw} \simeq \frac{h_0^2}{\num{5.6e-6}} \int_{f_\text{min}}^{f_\text{max}} d(\ln f) \ \Omega_\text{gw}(f) \ .
\end{equation}
The term $\Delta N_\text{eff}^\text{gw}$ represents the GW contribution to the effective number of relativistic species, $\Delta N_\text{eff}$, and is constrained by BBN and CMB probes to $\Delta N_\text{eff}^\text{gw} \lesssim 0.4$ at a $2 \sigma$ upper limit \cite{Planck:2018vyg, 2020ApJ...896...77H, ACT:2020gnv}. The frequencies $f_\text{min}$ and $f_\text{max}$ depend on the epoch of BBN and the reheating temperature $T_\text{rh}$, respectively. Only modes within the horizon at the time of BBN can contribute to $\Delta N_\text{eff}$, meaning $f_\text{min}$ will be determined by the size of the horizon at the beginning of BBN. This scale corresponds to $10^{-12} \ \text{Hz}$, but in actuality, modes will need to begin oscillating to contribute to the radiation energy density. As such, we will set $f_\text{min} = 10^{-10} \ \text{Hz}$ \cite{Cabass:2015jwe}. Assuming instantaneous reheating after inflation ends, the upper limit $f_\text{max}$ is roughly proportional to the reheating temperature $T_\text{rh}$. We benchmark this proportionality with GUT-scale reheating, where $T_\text{rh} \simeq 10^{15} \ \text{GeV}$ corresponds to $f_\text{max} \simeq 10^8 \ \text{Hz}$.

\section{Data and Methods} \label{data and methods}
As mentioned above, the NANOGrav collaboration reported evidence for a stochastic signal that was correlated among 67 pulsars from a 15-year pulsar-timing dataset \cite{NANOGrav:2023gor}. 
Importantly, the correlations follow the Hellings-Downs pattern expected for a stochastic GW background, pointing to a true GW origin for the source \cite{Hellings:1983fr}.\footnote{{ Several measurement uncertainties are present, including pulsar variance (due to pulsar sampling) and cosmic variance (due to Gaussian signals) \cite{Bernardo:2022xzl}}}. 
The NANOGrav collaboration performed extensive analysis on this 15-year pulsar-timing dataset and constructed a joint posterior probability distribution of the intrinsic-noise amplitude $A$ and the spectral index $\gamma$ in a Hellings-Downs power law model \cite{NANOGrav:2023gor}. This posterior distribution, shown in Fig. 1b (the blue contours using $f_\text{ref} = 1 \ \text{yr}^{-1}$) of Ref. \cite{NANOGrav:2023gor}, will be used as the primary signal for our PTA data.

To constrain the region of inflationary parameter space ($\log_{10} r$ vs. $n_T$ space) corresponding to the NANOGrav posterior, a grid scan could be performed using Eq. \ref{eq: n_T(gamma)}-\ref{eq: A(n_T, r)}. In fact, this method was initially utilized to identify the general characteristics of the contours in inflationary parameter space. However, we choose a more versatile technique and perform a Bayesian analysis by using a Markov Chain Monte Carlo (MCMC) algorithm to map out the region of inflationary parameter space. We'll use the Metropolis-Hastings algorithm to dictate the sampling of the NANOGrav posterior, and we'll evaluate the convergence of the MCMC chains using the Gelman-Rubin parameter $R - 1$, requiring $R - 1 < 10^{-3}$ for a run \cite{gelman10.1214/ss/1177011136}. 

\begin{figure}[ht]
\centering
\includegraphics[width=\linewidth]{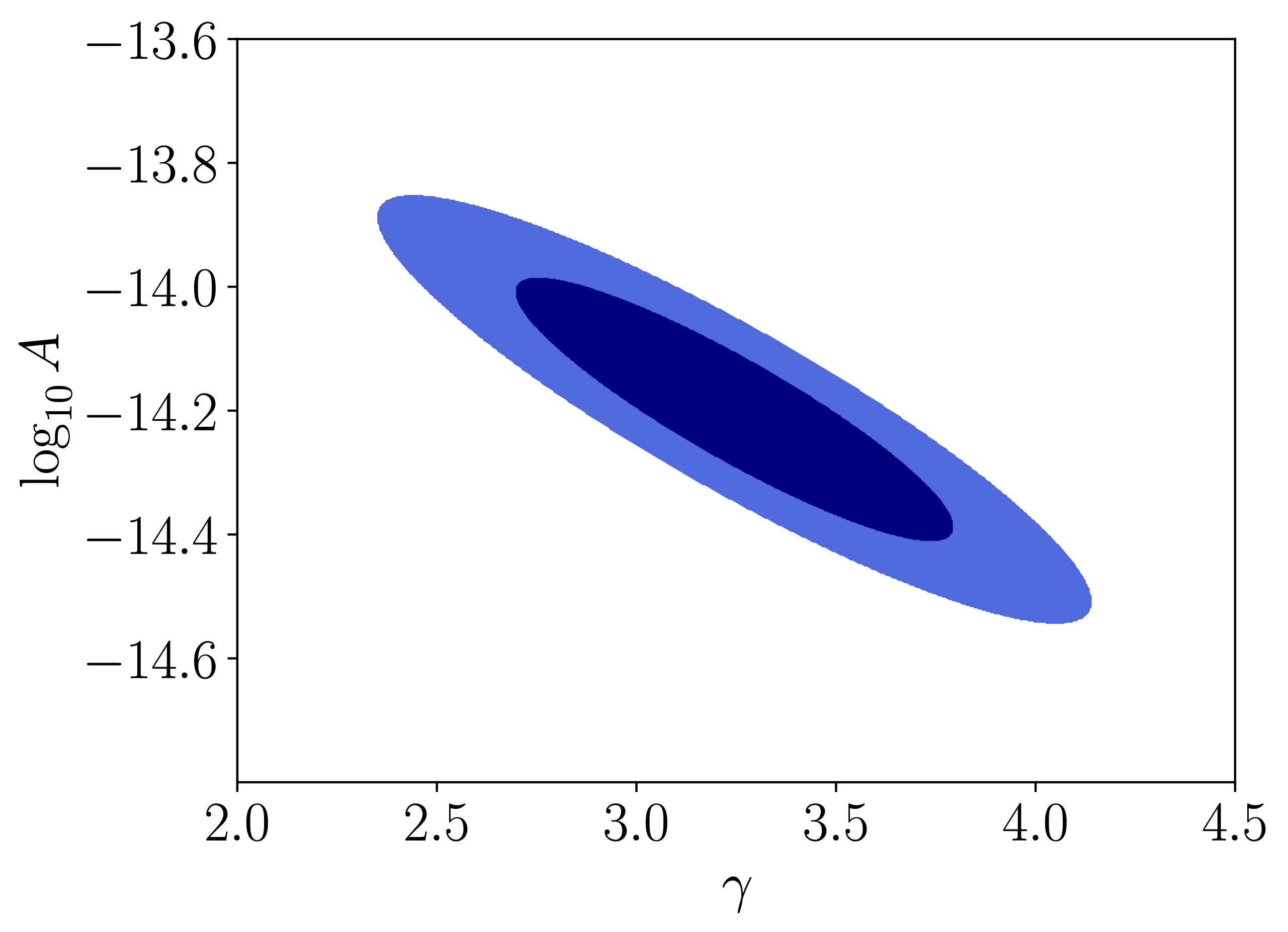}
\caption{The 2D joint posterior probability distribution for the logarithm of the intrinsic-noise amplitude $\log_{10}A$ and the spectral index $\gamma$, using a multivariate Gaussian distribution to approximate the NANOGrav 15-year posterior. Comparing this distribution with Fig. 1b of Ref. \cite{NANOGrav:2023gor} shows excellent agreement between the two, confirming the validity of this approximation. }
 \label{fig: gaussian}
\end{figure}

The NANOGrav posterior will serve as the likelihood for the MCMC run. We find that the NANOGrav posterior is well approximated as a multivariate Gaussian distribution over $\log_{10} A$ and $\gamma$. This Gaussian can be described via its mean vector $\boldsymbol{\mu}_N$ and its covariance matrix $\boldsymbol{\Sigma}_N$:
\begin{align}
\boldsymbol{\mu}_N & \simeq [3.25, -14.19] 
\\ \boldsymbol{\Sigma}_N & \simeq \begin{bmatrix}
0.129 & -0.045 \\
-0.045 & 0.0191
\end{bmatrix} \ .
\end{align}
\\ \\
The joint posterior distribution from this approximation is show in Fig. \ref{fig: gaussian}. Using this distribution, the log-likelihood for the MCMC runs is:
\begin{equation} \label{eq: likelihood}
    \ln \mathcal{L}(\boldsymbol{\theta}) = -\frac{(\boldsymbol{x}(\boldsymbol{\theta})-\boldsymbol{\mu}_{N})^T\boldsymbol{\Sigma}_{N}^{-1}(\boldsymbol{x}(\boldsymbol{\theta})-\boldsymbol{\mu}_{N})}{2} \ ,
\end{equation} 
where $\boldsymbol{x}(\boldsymbol{\theta}) \equiv [\log_{10}A(\boldsymbol{\theta}), \gamma(\boldsymbol{\theta})]$ is a vector of the dependent parameters, $\boldsymbol{\theta}$ is a vector of the independent parameters, which we take to be $\boldsymbol{\theta} = [r, n_T]$, and $^T$ denotes the transpose operation. The functions $\log_{10}A(\boldsymbol{\theta})$ and $\gamma(\boldsymbol{\theta})$ used to calculate $\boldsymbol{x}(\boldsymbol{\theta})$ are given by Eq. \ref{eq: A(n_T, r)} and Eq. \ref{eq: n_T(gamma)}, respectively. Technically, $\boldsymbol{\theta}$ should include all cosmological parameters relevant to an inflationary GW, i.e. $H_0, A_s, \Omega_m, \Omega_r, n_T$ and $r$, but we simply fix all but $n_T$ and $r$ to their \textit{Planck} 2018 values \cite{Planck:2018vyg}. 

This likelihood in Eq. \ref{eq: likelihood} is supplied to the cosmological sampler \texttt{MontePython}, which we utilize to perform the MCMC runs \cite{Brinckmann:2018cvx, Audren:2012wb}. We also choose to work with $\log_{10}r$ instead of $r$, as the magnitude of $r$ is still unknown. We impose flat, wide priors on both $n_T$ and $\log_{10}r$, and impose a hard prior of $\log_{10}r < -1.44$ corresponding to the $2\sigma$ constraint $r < 0.036$ from analysis of \textit{Planck}, WMAP, BICEP3, and \textit{Keck} data \cite{BICEP:2021xfz}.

\begin{figure*}[ht]
\centering
\includegraphics[width=0.8\textwidth]{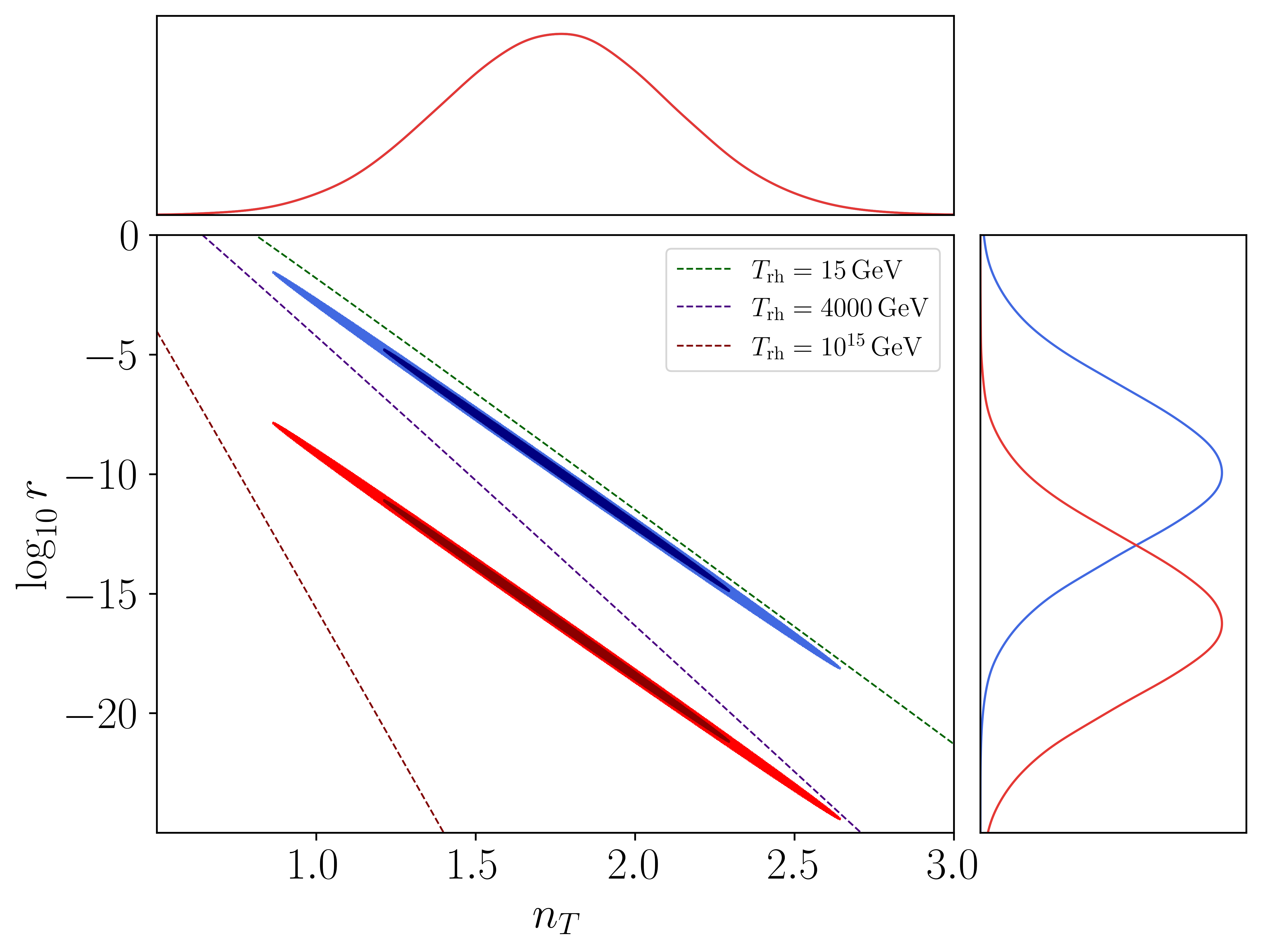}
\caption{{The 2D joint posterior probability distribution for the logarithm of the tensor-to-scalar ratio $r$ and the tensor spectral index $n_T$, generated by mapping the joint posterior probability distribution from Fig. \ref{fig: gaussian} via MCMC runs. Following Fig. \ref{fig: mode evolution}, the blue contours are for the GR model and the red contours are for the MG model, which inherently depend on the mass $m = 10^9 \cdot H_0$ and the cutoff time $\tau_m = 5\tau_{\rm eq}$ (see Section \ref{transfer function}). The dashed lines are produced by fixing $\Delta N_\text{eff}^\text{gw} = 0.4$ in Eq. \ref{eq: BBN integral}, and setting the integral upper bound $f_\text{max}$ to the a value corresponding to a particular reheating temperature $T_\text{rh}$. The region to the left of a dashed line is acceptable under that value of the reheating temperature.}}
 \label{fig: contours}
\end{figure*}

\section{Results} \label{results}
We perform MCMC runs as described in the previous section to identify the region of inflationary parameter space corresponding to the NANOGrav joint $A$-$\gamma$ posterior probability distribution using the model of GR and the model of MG. Specifically, we set $s_1 = 4.16$ for the GR case and 
$s_1 = 1.3 \cdot 10^6$
in the MG case in Eq. \ref{eq: A(n_T, r)}, which affects the likelihood in Eq. \ref{eq: likelihood} as described in Section \ref{data and methods}. The results of these MCMC runs are shown in Fig. \ref{fig: contours}, where the color scheme matches that of Fig. \ref{fig: mode evolution}: i.e. blue for the GR model and red for the MG model. 

Looking at the $1 \sigma$ contours, we find that:
\begin{align}
    n_T &= 1.75 \pm 0.54 
    \\ \log_{10}r &= -9.83 \pm 5.05 \ ,
\end{align}
for the GR model, while:
\begin{align}
    n_T &= 1.75 \pm 0.54 
    \\ \log_{10}r &= -16.14 \pm 5.05 \ ,
\end{align}
for the MG model. Observe that the mean of $\log_{10}r$ is lower in the MG model, as a result of the different value of $s_1$. The $1\sigma$ and $2\sigma$ contours for the GR and MG models both show a negative correlation between $n_T$ and $\log_{10}r$, which is expected. Recalling the form of Eq. \ref{eq: prim power spectrum}, we see that higher values of $n_T$ can be compensated by lower values of $r$, and vice versa. This behavior is independent of the choice of $s_1$, and thus appears for both the GR and MG models. The relationship between $\gamma$ and $n_T$, given by Eq. \ref{eq: n_T(gamma)}, is also independent of $s_1$, which explains why the range of values for $n_T$ is shared for both models.

Since the range of $n_T$ for both models is much greater than zero, we characterize the spectrum as strongly blue; this strongly blue spectrum presents challenges for inflationary justification. First, observe that this spectrum violates the consistency relation discussed in Section \ref{model setup}, which requires a slightly red spectrum. This relation only applied to the simple single-field slow-roll inflationary model, meaning other models may suffice. Second, reconsider Eq. \ref{eq: eos n_T} and observe that $n_T = 1.75$ would require $w = -\frac{17}{3}$, which is difficult to reach theoretically due to being deep within the phantom regime. However, previous work has investigated how double-field inflation \cite{Adams:1990ds, Ashoorioon:2015hya, Ashoorioon:2022raz, Tsujikawa:2002qx}, axion inflation \cite{Niu:2023bsr, McAllister:2008hb, Flauger:2009ab}, or quintessential inflation \cite{Das:2023nmm, Peebles:1998qn, Sami:2004xk} could produce a sufficiently blue spectrum to match the NANOGrav signal. Other mechanisms within previous models can also be adequate, such as scalar-induced GWs \cite{Frosina:2023nxu} or the removal of the Bunch-Davies initial condition \cite{Choudhury:2023kam}, both of which utilize single-field inflation. As such, the strongly blue spectrum can be justified, but with some difficulty.

We now turn our attention to the inflationary parameter $\log_{10}r$ in Fig. \ref{fig: contours}, for which we see that the range of values of $\log_{10}r$ for the MG model is lower than that of the GR model. The significance of this comes from its effect on the fixed reheating temperature isocontours shown as dashed lines in Fig. \ref{fig: contours}. These lines come from the BBN bound in Eq. \ref{eq: BBN integral}, using $\Delta N_\text{eff}^\text{gw} = 0.4$ (see Section \ref{transfer function}), and setting $f_\text{max}$ based on the chosen value of the reheating temperature $T_\text{rh}$. The region to the left of a particular dashed line represents the region of $n_T$-$r$ parameter space that agrees with the chosen reheating temperature $T_\text{rh}$ according to the BBN integral from Eq. \ref{eq: BBN integral}. The three reheating temperatures considered for Fig. \ref{fig: contours} are $T_\text{rh} \simeq 15 \ \text{GeV}$, $T_\text{rh} \simeq 4000 \ \text{GeV}$, and $T_\text{rh} \simeq 10^{15} \ \text{GeV}$. The former two temperatures have been chosen to constrain the GR and MG models respectively, while the latter temperature is the GUT-scale reheating temperature, used in Section \ref{transfer function} to calibrate the proportionality between $f_\text{max}$ and $T_\text{rh}$. Importantly, the reheating temperature for the MG model is higher than reheating temperature for the GR model, due to the lower mean value of $\log_{10}r$ for the MG model. These reheating scales are quite low, but not excluded from possibility, as the reheating temperature can be as low as $5 \ \text{MeV}$ according to previous probes \cite{deSalas:2015glj, Kawasaki:2000en, Hasegawa:2019jsa}. However, similar to a bluer spectrum in $n_T$, a lower reheating temperature is harder to motivate theoretically. Hence, although both MG and GR models are technically viable, the MG model is less cumbersome to motivate due to its higher reheating temperature. 

Overall, an inflationary interpretation of the NANOGrav signal remains challenging, as shown in other works \cite{Vagnozzi:2020gtf, Vagnozzi:2023lwo, Odintsov:2023weg, Choudhury:2023hfm, Choudhury:2023kam, Niu:2023bsr, Jiang:2023gfe}. The inclusion of a MG model relaxes the difficulty of explaining the low reheating temperature, but the strongly blue spectrum is still tough to motivate. The possibility of an inflationary interpretation in the GR model is poor due to constraints from both the low reheating temperature and the blue spectrum, while the MG model's interpretation is slightly better due to its higher reheating temperature.

\section{Conclusions} \label{conclusions}
In this work, we've considered how the recently released NANOGrav 15-year dataset could be interpreted as an inflationary GW background by using MTMG with a step function mass instead of standard GR. Using a modified EoM, a result of MTMG, we've calculated a transfer function that describes the evolution of GW modes across the universe's history. This transfer function was then used to derive a new GW spectral energy density, which would serve at the connecting point between the inflationary theory and the PTA signal from NANOGrav. As such, we were able to map the NANOGrav $A$-$\gamma$ joint posterior probability distribution onto the $n_T$-$r$ inflationary parameter space, by using a MCMC scan. The relevant regions of inflationary parameter space are shown in Fig. \ref{fig: contours}, which shows $1\sigma$ and $2\sigma$ contours for both the GR and MG models. Analyzing this figure, we find that an inflationary interpretation would require a strongly blue spectrum, with a relatively low reheating temperature. Though both of these constraints are theoretically viable, motivating both of them remains challenging. Importantly, the MG model permits a higher reheating temperature than the GR model, slightly relaxing the difficulty of an inflationary interpretation in the MG case.

As discussed in Section \ref{transfer function}, a specific graviton mass and cutoff time were selected in our work prior to calculating the transfer function for MG via numerical integration. Future work may consider different ranges for graviton mass, or later cutoff times, to evaluate how each of these affect the transfer function. In particular, it may be fruitful to determine what ranges for the graviton mass and cutoff time within previously studied bounds would lead to the most tenable inflationary interpretation. 

\subsection{Data Availability}
The NANOGrav 15-year Data Set used in this paper is available from NANOGrav \cite{NANOGrav:2023}. Source code required to reproduce the analysis and figures in this paper are available in the GitHub repository \cite{github}.
\subsection{Acknowledgements}
We'd like to thank Murman Gurgenidze and Sayan Mandal for helpful discussions, as well as Chris Choi and Jacob Magallanes for useful cross-checks of GW evolution in the MTMG model. Furthermore, we thank Sunny Vagnozzi for assistance with the MCMC runs. TK acknowledges partial support from the NASA Astrophysics Theory Program (ATP) Award 80NSSC22K0825 and National Science Foundation (NSF) award AST2408411.



\bibliographystyle{apsrev4-2_edited}
\bibliography{refs}
\end{document}